\documentclass[letterpaper,10pt,conference]{IEEEtran}

\usepackage{copyright}

\usepackage[utf8]{inputenc}
\usepackage[english]{babel}
\usepackage{subcaption}

\usepackage{graphicx}
\usepackage{cite}

\usepackage[]{quoting}

\usepackage[binary-units]{siunitx}
\usepackage[acronym]{glossaries}
\glsdisablehyper
\newacronym{jlr}{JLR}{Jaguar Land Rover}
\newacronym{fmcw}{FMCW}{Frequency-Modulated Continuous-Wave}
\newacronym{mmw}{MMW}{Millimetre-Wave}
\newacronym{lidar}{LiDAR}{Light Detection and Ranging}
\newacronym{gps}{GPS}{Global Positioning System}
\newacronym{6dof}{6DoF}{six degree-of-freedom}
\newacronym{av}{AV}{autonomous vehicle}
\newacronym{ad}{AD}{Autonomous Driving}
\newacronym{aaip}{AAIP}{Assuring Autonomy International Programme}
\newacronym{rtk}{RTK}{Real-Time Kinematic}
\newacronym{gnss}{GNSS}{Global Navigation Satellite System}
\newacronym{ai}{AI}{Artificial Intelligence}
\newacronym{ddn}{DDN}{Distilled Dropout Network}

\usepackage{xcolor}

\newcommand{\reviewA}[1] {{\color{orange} {}}}
\newcommand{\reviewB}[1] {{\color{yellow} {}}}

\usepackage{hyperref}
\usepackage{url}
\newcommand\rurl[1]{%
  \href{http://#1}{\nolinkurl{#1}}%
}

\usepackage{cleveref}
\crefname{table}{Table}{Tables}
\crefname{figure}{Figure}{Figures}
\crefname{section}{Section}{Sections}

\IEEEoverridecommandlockouts

\begin{document}

%------------------------------------------------------------------
\title{Sense--Assess--eXplain (SAX): \\Building Trust in Autonomous Vehicles in Challenging Real-World Driving Scenarios}
\author{Matthew Gadd$^{1,*}$, Daniele De Martini$^{1,*}$, Letizia Marchegiani$^2$, Paul Newman$^1$, and Lars Kunze$^1$
\\
$^1$Oxford Robotics Institute, Dept. Engineering Science, University of Oxford, UK.\\\texttt{\{mattgadd,daniele,pnewman,lars\}@robots.ox.ac.uk}
\\
$^2$Automation and Control, Dept. Electronic Systems, Aalborg University, DK.\\\texttt{lm@es.aau.dk}
\thanks{$^*$ Matthew Gadd and Daniele De Martini contributed equally to this work.}
}
\maketitle
%------------------------------------------------------------------

\copyrightnotice

%------------------------------------------------------------------
\begin{abstract}
This paper discusses ongoing work in demonstrating research in mobile autonomy  in challenging driving scenarios. In our approach, we address fundamental technical issues to overcome critical barriers to assurance and regulation for large-scale deployments of autonomous systems.
To this end, we present how we build robots that (1) can robustly sense and interpret their environment using traditional as well as unconventional sensors; (2) can assess their own capabilities; and (3), vitally in the purpose of assurance and trust, can provide causal explanations of their interpretations and assessments.
As it is essential that robots are safe and trusted, we design, develop, and demonstrate fundamental technologies in real-world applications to overcome critical barriers which impede the current deployment of robots in economically and socially important areas.
Finally, we describe ongoing work in the collection of an unusual, rare, and highly valuable dataset.
\end{abstract}

%------------------------------------------------------------------
\section{Introduction}
%------------------------------------------------------------------

\reviewA{
The paper is heavily focused on describing future work. I
did appreciate the amount of work that was put in the
making of the article; however, I think that it is too
early for technical publication.
}

\reviewB{
The paper is not difficult to read but there is lack of
correlation among the sentences and paragraphs. Also, it
was better the authors start with some fundamental
definitions to clear included aspects and their goal. 
The authors presented a good problem discussion in the
introduction, but still more information needed to be
provided to better contextualize the work and provide a
better rationale for the approach taken. 
Meanwhile, it will be better to present more related works
and background about the topic. I think the related work
needed additional information, including a discussion of
the importance of the context.
Generally, I agree that the authors did a good job of
describing the issue and representation that it is timely,
and interesting. I also feel that work appears to be mostly
sound, and appreciated the range of study activities
undertaken. 
}

The perception and navigation capabilities of autonomous vehicles have been tremendously improved over the past decade.
However, to increase the level of trust in autonomy in driving scenarios and to assure safety during operation, a range of open challenges need to be addressed. 
These challenges include:
\begin{enumerate}
\item \textbf{robust perception} of real-world environments under changing weather conditions, 
\item \textbf{introspection} and \textbf{assessment} of perception and navigation processes, and the 
\item \textbf{semantic interpretation} and \textbf{explanation} of scenes as well as the vehicle's performance.
\end{enumerate}

\begin{figure}
    \centering
    \includegraphics[width=0.8\columnwidth]{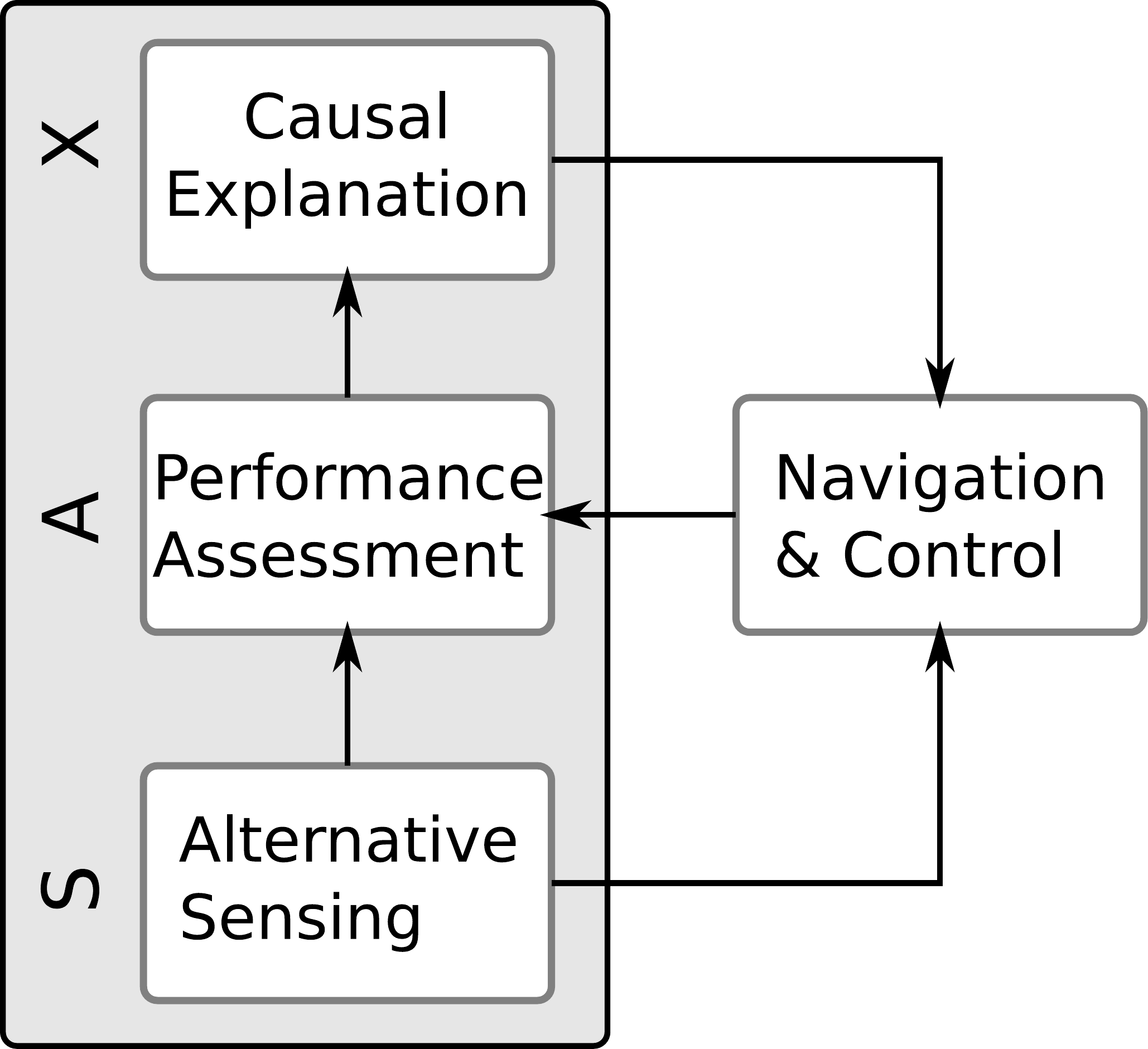}
    \caption{
    An illustration of our Sense-Assess-eXplain (SAX) interpretation of trust and assurance of \acrlong{av} operations.
    \emph{Alternative Sensing} is about the perception of the environment using unconventional sensing methods.
    \emph{Performance Assessment} then develops methods necessary to assess perception systems as well as navigation and control aspects.
    \emph{Causal Explanation} then anchors these representations in situated driving scenarios and develops methods for generating causal explanations.
    \emph{Navigation and Control} is not the focus of our research agenda in this project.
    However, as it is an essential component of the overall integrated system we will leverage existing resources in this area.}
    \label{fig:wps}
    \vspace{-0.4cm}
\end{figure}

In this work, as illustrated in~\cref{fig:wps}, we address the missing link between unconventional sensing modalities and environmental performance assessment in real-world scenarios, in conjunction with and coupled to explainability.
Our work is built around the following concrete objectives with clear measurements of success through which we aim to impact the way in which we trust and assure autonomy of autonomous vehicles:

\begin{enumerate}
  \item[\textbf{O1:}] To robustly and reliably sense and interpret the environment in severe and changing weather conditions, overcoming the limitations of classic sensing modalities;
  \item[\textbf{O2:}] To continuously assess and optimise the performance of perception as well as navigation methods;
  \item[\textbf{O3:}] To demonstrate a system capable of explaining in non-engineering and human terms what a robot/vehicle has seen and how it has influenced its decision making. 
\end{enumerate}

This paper is organised as follows.
\cref{sec:related_work} discusses related projects being undertaken in the community.
\cref{sec:motivation} describes a thought experiment which we use to suggest a framing of the open issues in~\cref{sec:paradigm}.
Our particular approach to these challenges is described in~\cref{sec:approach}.
\cref{sec:findings} describes our initial findings, a valuable dataset that we are collecting, and the schedule for investigation going forwards.
\cref{sec:discussion,sec:conclusion} discuss our contribution and future avenues for investigation.

%------------------------------------------------------------------
\section{Context for this Study}
\label{sec:related_work}
%------------------------------------------------------------------

% aaip
This project is firmly embedded within and aligned with the outcomes of the \gls{aaip}\footnote{\url{https://www.york.ac.uk/assuring-autonomy/}} -- addressing global challenges in assuring the safety of robotics and autonomous systems.
Related \gls{aaip} work which may be of interest to the reader includes~\cite{porter2018moral,mcdermid2019towards,burton2019confidence,bloomfield2019disruptive,burton2020mind,menon2019safety}.

% evsav
In the broader community and as exhibited in focused sessions~\cite{elli2nd,mccall20161st} we find that the works of~\cite{salay2019safety,bagschik2016identification,gerdes2019designing,pechberti2010design,cui2018development,johnson2019colliding,julian2020validation,zhan2019interaction,bagschik2018system,mohan2019method,naumann2018generating,Koopman2019AutonomousVM,wang2019overcoming,menzel2019functional} have a bearing on this research agenda.
In particular, works dealing with validation and proving safety~\cite{julian2020validation,Koopman2019AutonomousVM}, critical scenarios and hazardous events~\cite{bagschik2016identification,wang2019overcoming}, human values and comfort~\cite{gerdes2019designing,naumann2018generating}, human-understandable descriptions~\cite{menzel2019functional}, safety-oriented architectures~\cite{bagschik2018system,mohan2019method}, and mitigating hazardous events~\cite{johnson2019colliding} all related to our objectives.
Furthermore, from a robotics perspective, the role of perceptual components in safety systems~\cite{salay2019safety} as well as demonstrators~\cite{pechberti2010design}, simulators~\cite{cui2018development} and datasets~\cite{zhan2019interaction} for tackling these research challenges are all approaches that inspire the work presented.

In this work, we draw on our experience in robust navigation~\cite{churchill2013experience} and scene understanding~\cite{kunze2018reading} as demonstrated in trials that we have executed in challenging scenarios~\cite{oro58259,oro48661,BrianYeomansASTRA2017}.
Specifically, we continue to advocate the use of commercially promising but unusual sensing technology which is inherently robust to inclement weather and illumination\cite{MarchegianiTAROS2018,MarchegianiICRA2017,tang2020rsl,KidnappedRadarICRA2020}.

%------------------------------------------------------------------
\section{Motivation}
\label{sec:motivation}
%------------------------------------------------------------------

To illustrate the assurance paradigm we advocate in~\cref{sec:paradigm} and the approach we describe in~\cref{sec:approach} to solve the inherent issues this paradigm captures, let us consider the following concrete example scenario:

\begin{quoting}[leftmargin=10pt,rightmargin=1pt]
While driving off-road, a vehicle enters a region it has not traversed before.
It leverages external services to retrieve satellite images which provide a large-scale overview of the region ahead.
Based on these images, the vehicle creates a map, performs a semantic segmentation of driving surfaces, and plans a route through them according to their traversability.
While following the route, radar and audio sensor measurements are used to refine and update the surface segmentation in the map -- in the long-range and short-range respectively.
The vehicle will also explain to a human driver \textbf{what} route was taken, and \textbf{why}: \emph{``The vehicle will take a route over an area of gravel.
The route is slightly longer than the direct route, as there is an non-traversable body of water in the direction of the goal.''}

Next, consider that as it starts to rain heavily, the vehicle notices a drop in its localisation performance using its cameras.
Due to the change of weather conditions -- which is also detected through audio (change of surface properties) and confirmed by external weather services -- the vehicle seamlessly adapts its localisation system from camera to radar and reduces its velocity.
Although this process happens in the background, the vehicle can \textbf{explain} the cause of its decision to the human driver: \emph{``Due to the heavy rain and slippery surface conditions, the vehicle has reduced its speed.''}
An explanation to a developer and/or system auditor will provide more technical details: \emph{``Due to a 5\% drop in localisation performance using the camera the vehicle switched to a localisation method using radar.''}
\end{quoting}

For full capability in these scenarios in a fashion that is understandable and comfortable for a human occupant or auditor, the autonomous vehicle must:
\begin{enumerate}
    \item be able to robustly sense and understand its environment,
    \item have a good understanding of how well its various (perceptual or otherwise) subsystems are performing for the task at hand in the current driving conditions, and
    \item relay this information to a human occupant/auditor in a rational and digestible format.
\end{enumerate}

We capture these three aspects in~\cref{sec:paradigm} which frames our research agenda.
\Cref{sec:approach} describes our proposed approach in order to answer these research questions.
We acknowledge alternative approaches, however, and hope that the paradigm itself finds use in the broader community.

%------------------------------------------------------------------
\section{The Sense-Assess-eXplain (SAX) Paradigm}
\label{sec:paradigm}
%------------------------------------------------------------------

The proposed approach to trust and assurance addresses these challenges using a paradigm called \textbf{Sense-Assess-eXplain (SAX)} which comprises three complementary strands of research:        
To summarise, we \emph{explain} at different levels of abstraction what we have \emph{sensed} and \emph{assessed}. 
This means that, while driving, the vehicle is able to explain what it has perceived and how this has influenced its own decision making. 
Moreover, the vehicle will be able to explain how its performance depends on the current and predicted environmental conditions.

\subsection{Sense} 
We shall \emph{sense} the world through a set of unconventional but complementary sensors -- including \gls{fmcw} scanning radar and acoustic sensors -- that will allow us to perceive and interpret the environment in novel ways beyond the current state-of-the-art.
These alternative sensing methods will allow us to make robust perceptions where traditional sensing modalities might fail under severe weather conditions.
We take the view that these new additional modalities, so rarely used, offer both an axis of assurance and validation viz-a-viz conventional established techniques \emph{and} an expansion of the operating envelope.
In particular, we focus on the perception of driving surfaces in on-road and off-road scenarios under various weather (including torrential rain and snow) and lighting conditions using radar as well as the interpretation of complex, unstructured environments using auditory sensing.
Finally, to increase the vehicle's environmental awareness we sense the environment through a set of available data services such as rain radar (from weather services) and satellite imagery. 

\subsection{Assess} 
We \emph{assess} both environmental conditions (a priori) and the vehicle's perceptual as well as navigational performance in order to increase its environmental awareness and to adapt its behaviour accordingly.
To this end, we shall continuously assess the performance and effectiveness of perception and navigation methods and adapt them if necessary.    

\subsection{eXplain}

This ``explanatory'' thread lies at the heart of our assurance and trust research agenda.
We feel it is sorely lacking in much of the \gls{av} research endeavour where, with good reason, black boxes abound; however, the narrative is clearly changing.
We have to be able to offer users pathways to trust and assurance of \emph{what} machines are doing and \emph{why}.
They have to be able to explain, almost justify, what they are doing, what they see, what it means, and what they plan to do next and why.
The inputs to this explanatory process are, of course, the \emph{Sense} and \emph{Assess} threads, and their value is exponentiated when they are used to give humans common-sense everyday explanations of intended action and perception.
We posit that human users demand this as a precursor to trust, and we know that commercial insurers will require it.

\begin{figure}
    \centering
    \includegraphics[width=\columnwidth]{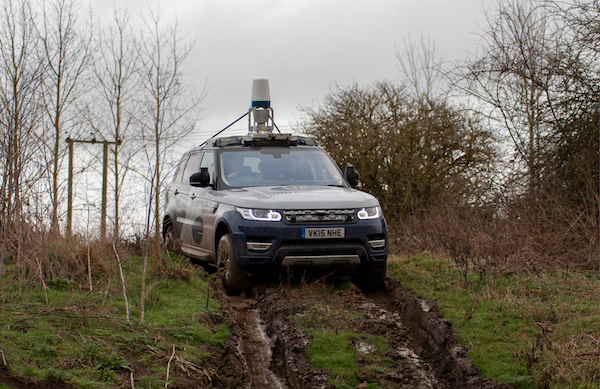}
    \caption{Our primary vehicle, a \acrlong{jlr}, while traversing an off-road scenario during one of the data-collection trials.\label{fig:jlr}}
    \vspace{-0.6cm}
\end{figure}

%------------------------------------------------------------------
\section{The Sense-Assess-eXplain (SAX) Methodology}
\label{sec:approach}
%------------------------------------------------------------------

\begin{figure}
    \centering
    \includegraphics[width=\columnwidth]{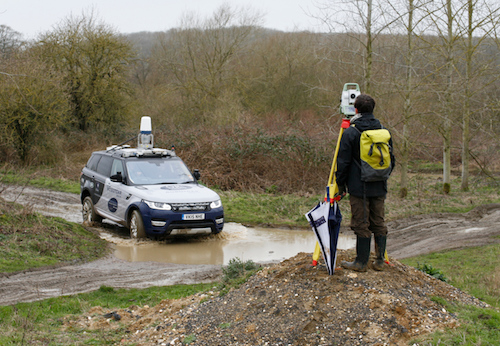}
    \caption{The Leica laser tracker used for ground-truth recording during one of the off-road trials.\label{fig:gadd_gt}}
    \vspace{-0.6cm}
\end{figure}

In this project we build on our perception, mapping, and localisation capabilities -- the ideal substrate to perceive challenging environments.
In our approach for sensing, assessing, and explaining the environment we harness the power of deep learning, while we utilise structure, priors, and models to guide the learning process.
By combining deep learning with \gls{ai} reasoning methods and structure we can overcome some of the critical barriers for assuring autonomy.
For example, a vehicle will be able to provide detailed causal explanations of its decisions at different levels of abstractions for different stakeholders. 
As discussed in~\cref{sec:findings}, our approach will be validated in complex, real-world driving scenarios using the \gls{jlr} platform shown in~\cref{fig:jlr,fig:gadd_gt}.

%------------------------------------------------------------------
\subsection{Alternative Sensing}

In the last decades, many advances have been made in \gls{av} navigation and localisation.
Nevertheless, these are still open problems, especially when \glspl{av} are deployed into the real world, exposing challenges that are hardly predictable in the laboratories, particularly in the perception of the environment.
Harsh weather and lighting conditions in particular pose non-trivial challenges to \gls{av} development, above all with the usage of traditional sensing systems, as cameras and \gls{lidar}.

Since all autonomous tasks are built on top of environment perception, the availability of robust sensing information, as well as algorithms and techniques to interpret it, is crucial for all robotic platforms.
The objective of this line of work is therefore to investigate the exploitation of uncommon sensing modalities and configurations, such as scanning radars and audio, and external weather and map services, for the vehicle's motion estimation and surface classification.

Furthermore, new techniques such as deep learning can be extremely effective tools to model the data streams coming from such sensors.
For those models to be robust and accurate, though, a thorough and consistent dataset, representing a variety of experiences and conditions, is necessary.

Work in this area includes:
\begin{enumerate}
    \item Radar-based motion estimation and localisation,
    \item Auditory sensing,
    \item Leveraging external services (such as satellite imagery),
    \item Multi-modal terrain classification, and
    \item Data collection.
\end{enumerate}

\begin{figure*}
    \centering
    \begin{subfigure}{0.32\textwidth}
    \centering
    \includegraphics[width=\textwidth]{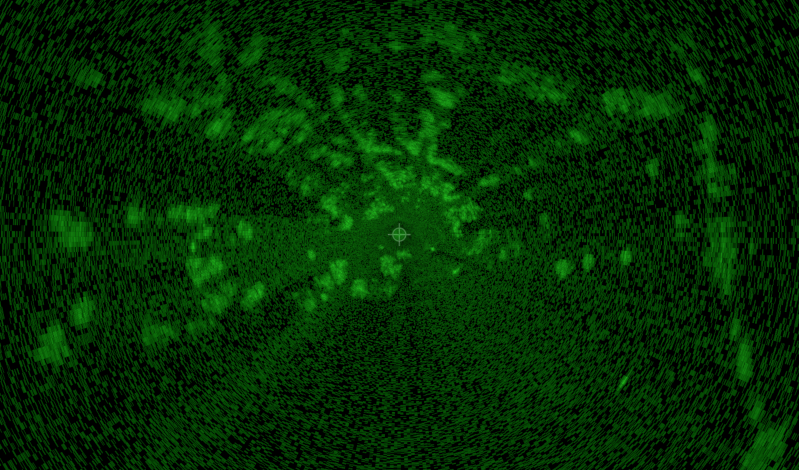}
    \caption{}
    \end{subfigure}
    \begin{subfigure}{0.32\textwidth}
    \centering
    \includegraphics[width=\textwidth]{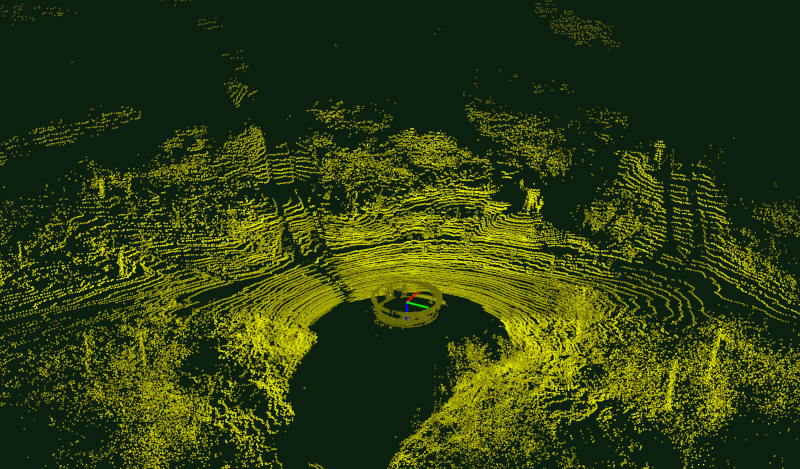}
    \caption{}
    \end{subfigure}
    \begin{subfigure}{0.32\textwidth}
    \centering
    \includegraphics[width=\textwidth]{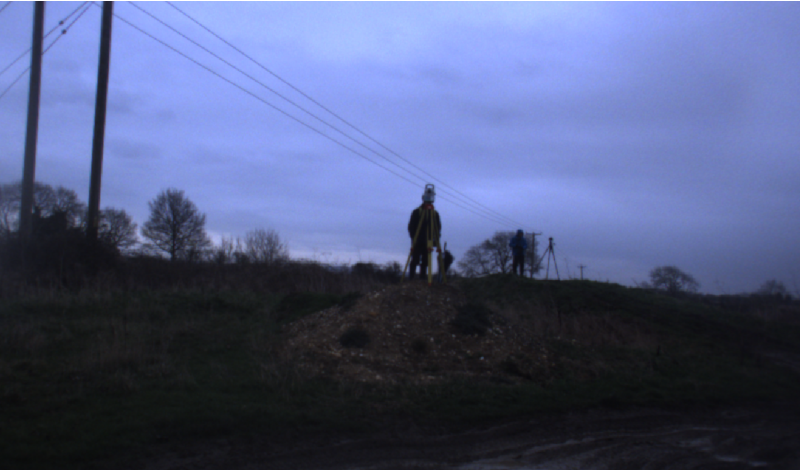}
    \caption{}
    \end{subfigure}
    
    \begin{subfigure}{0.32\textwidth}
    \centering
    \includegraphics[width=\textwidth]{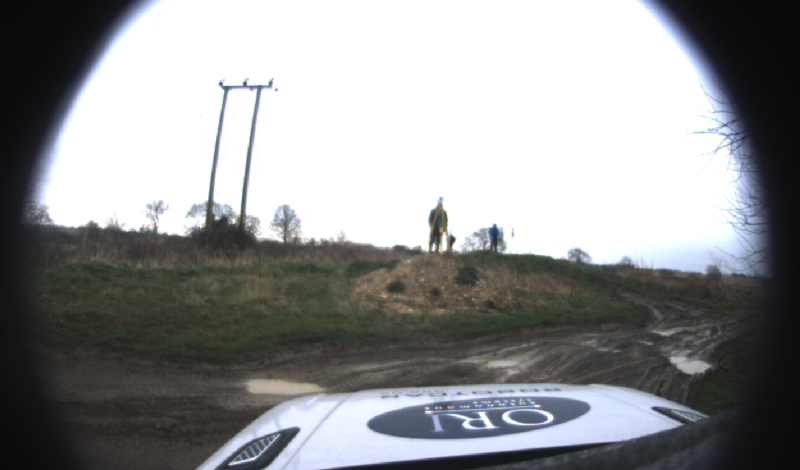}
    \caption{}
    \end{subfigure}
    \begin{subfigure}{0.32\textwidth}
    \centering
    \includegraphics[width=\textwidth]{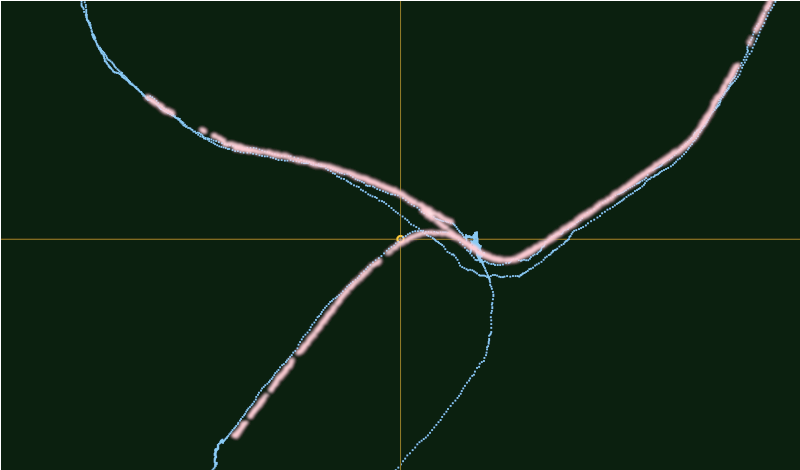}
    \caption{}
    \end{subfigure}
    \begin{subfigure}{0.32\textwidth}
    \centering
    \includegraphics[width=\columnwidth]{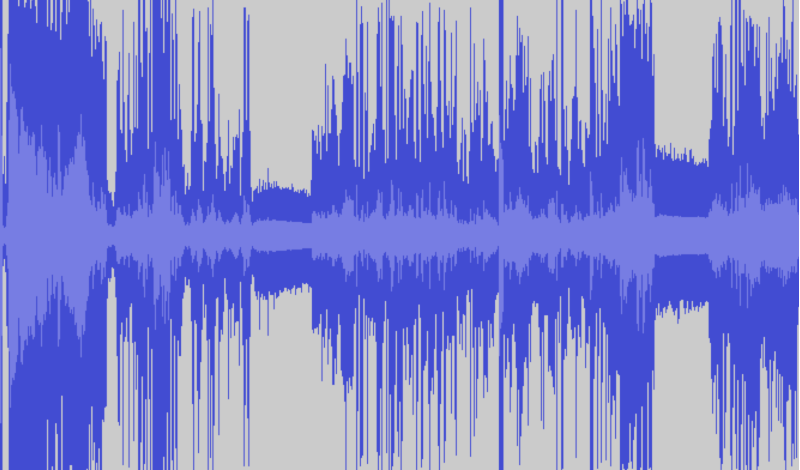}
    \caption{}
    \end{subfigure}

    \caption{Samples of sensor streams taken at the same time as in~\cref{fig:gadd_gt}. (a) shows a radar scan in its cartesian form; (b) shows 3D LiDAR data; (c) and (d) show the camera data taken by the from stereo and mono cameras respectively; (e) shows the manually-overlaid \gls{gps} and ground-truth data -- see~\cref{sec:findings}; (f) is the audio stream from the microphones in the wheel archs.\label{fig:sensor_data}}
    \vspace{-0.5cm}
\end{figure*}

\cref{fig:sensor_data} shows exemplar records taken from some of the sensors that our platform shown in~\cref{fig:jlr} is equipped with.

%------------------------------------------------------------------
\subsection{Performance Assessment}

Predicting the likely performance of a robotic sub-system based on past experience in the same workspace is applicable to both navigation~\cite{churchill2013experience} and perception~\cite{hawke2016wrong}.
In this context we will further equip \glspl{av} with accurate situational awareness for safe autonomous operation in complex environments.

Here, we consider that some environments are less lenient than others to even small lateral or rotational deviations from a known trajectory, so that localisation can be lost if the taught trajectory is not followed within some tolerance.
Furthermore, predictive uncertainty estimates from standard neural networks are typically overconfident, often making them too unreliable to be deployed in real world applications.

Work in this area includes:
\begin{enumerate}
    \item Predicting localisation performance, and
    \item Estimating model confidence.
\end{enumerate}

We plan in this challenge to draw on works such as~\cite{guruau2016fit} to intelligently and seamlessly select the sensing modality which has the most support in the region of the world currently experienced and its environmental condition.

%------------------------------------------------------------------
\subsection{Causal Explanation}

It is of key importance that users, developers, and regulators understand what a robot is doing, what it did, what it intends to do, and why. 
\emph{Explanations} are identified by \gls{aaip} as one of the critical barriers to assurance and regulation. 
Regulators are already requiring some form of \emph{interpretability} and \emph{explainability}.

In this project we will build robots that semantically understand their environment and can provide causal explanations for their own decisions. 
Transparent and interpretable representations will enable developers to analyse the robot's behaviour and assure its safe autonomous operation.
Users will benefit from explanations by developing trust in autonomous systems.

Work in this area includes:

\begin{enumerate}
    \item Scenario-based requirement analysis,
    \item Semantic scene representation, and
    \item Learning and inference for causal explanation.
\end{enumerate}

Here, we will investigate a range of driving scenarios to understand what types of explanations are required by stakeholders.
In particular, we will focus on scenarios which involve different types of surfaces and changing weather conditions.

Additionally, we will extend our graph-based scene representation~\cite{kunze18scenegraph} to encode information of traffic participants and other semantic aspects of the environment including the type of surfaces as well as weather information in a well-defined language (ontology). 

%------------------------------------------------------------------
\subsection{Integration and Demonstration}

Our \gls{av} demonstrator is used in transportation tasks in real-world environments.
The vehicle will perform these tasks in on-road and off-road settings on a range of different terrains and under different weather conditions.

The overall aim is to demonstrate that a vehicle can explain its observations of environmental conditions (e.g. surface, weather) as well as its own performance.

To this end, we will adopt the following principles: 

\begin{itemize}
    \item The vehicle uses a set of uncommon, multi-modal sensors (incl. radar and acoustic sensors) to mitigate against failure of traditional sensors such as cameras in severe weather conditions, and
    \item The robot assesses its performance in perception and navigation tasks and detects anomalies that are outside of learnt confidence bounds.
\end{itemize}

We will address our aim by exploring and validating the following practices:

\begin{itemize}
    \item Use of real-world datasets to learn sensor models including confidence bounds, for representative environments, and
    \item Validation of learnt models through use in real driving scenarios using our \gls{jlr} RobotCar.
\end{itemize}

Work in this area includes:

\begin{enumerate}
    \item Dataset release, and
    \item Real-world demonstration.
\end{enumerate}

Thus, the demonstrator will deliver a method for explaining observations of the environment as well as its assessments of perception and navigation routines, a dataset of unconventional sensors, and a taxonomy of explanations.
The taxonomy will provide guidance for regulators and system developers for the accreditation of \glspl{av}.

%------------------------------------------------------------------
\section{Preliminary and Upcoming Outcomes}
\label{sec:findings}
%------------------------------------------------------------------

Our early investigation has been focused on robust and effective motion estimation~\cite{2019ICRA_aldera,2019ITSC_aldera}, localisation~\cite{KidnappedRadarICRA2020,gadd2020lookaroundyou}, and semantic scene understanding~\cite{kaul2020rssnet} using \gls{fmcw} scanning radar.
We have also produced work showing effective localisation between satellite imagery and radar in~\cite{tang2020rsl}, towards leveraging external services.
Finally, we have released the largest radar-focused \emph{urban} autonomy dataset collected to date~\cite{RadarRobotCarDatasetICRA2020}.

However, many urban datasets have been released in recent years~\cite{yu2018bdd100k,nuscenes2019,lyft2019}.
For this reason, we are in the process of collecting a new dataset with the intention of encouraging research into \emph{introspection} and \emph{explainbility} of \gls{ad} systems in totally under-investigated driving scenarios.
To this end, our focus for data capture revolves around \emph{unusual sensing modalities}, \emph{mixed driving surfaces}, and \emph{adverse weather conditions}.

Specifically, we are completing driving in rural and off-road England with our \textit{Jaguar Land Rover RobotCar}.
We choose not to perform this investigation with smaller robotic platforms such as~\cite{kyberd2019} as despite the reach of these platforms into interesting scenarios which are important to economic sectors such as agriculture they are not \emph{also} suitable urban driving.

Indeed, we are capturing data in a wide set of atmospheric and illumination conditions while traversing a variety of surfaces including highways, country lanes, mud paths, and cobblestone streets.
Our sensor suite is comprised of sensors traditionally exploited for \gls{ad} including cameras, LiDARS, and GPS/INS.
However, we also include sensors which are not traditionally exploited but show great promise including \gls{fmcw} radar and audio.
Furthermore, we provide synchronised streams from external services including satellite imagery and publicly available weather information.
Towards building trust in \glspl{av} in challenging real-world driving scenarios, we also plan to provide human-derived metadata tags on the actions taken by the driver at interesting occasions during each foray.

Commonly for autonomous vehicles operating outdoor, high precision \gls{gps} is a viable way to record a ground-truth signal for algorithms to be tested and verified~\cite{RCDRTKArXiv}; nevertheless, an \gls{rtk} solution is dependant on the availability of a \gls{gnss} base station in close proximity~\cite{rtk_range} and fairly affected by occlusions -- trees and buildings -- and longitude.
Leica Viva TS16\footnote{\rurl{leica-geosystems.com/products/total-stations/robotic-total-stations/leica-viva-ts16}} is an accurate instrument used for surveying and building construction.
Since TS16 has robotic mode, it is able to lock onto the specific target and track it regardless of the amount of distractions in the field.
We axploit this excellent ability to provide a ground truth for accuracy evaluations.

%------------------------------------------------------------------
\section{Open Issues}
\label{sec:discussion}
%------------------------------------------------------------------

The work proposed in this research agenda is in a nascent field, with many open issues.
In this section we briefly list import challenges in this area which is either pending or out of scope of the work proposed here -- as framed by the paradigm discussed in~\cref{sec:paradigm}.

\paragraph{Sense}
Multi-modal hazard identification, safe environment perception in open environments, and dealing with the unknown.

\paragraph{Assess}
Safety argumentation for machine learning based systems, probabilistic guarantees, self-perception, self-awareness, introspection, related approaches of other communities, and determination of environment perception performance.

\paragraph{eXplain}
Performance evaluation of explanations, metrics and benchmarks for risk and safety, abstraction-levels and hierarchies for explanations, human-machine interfaces, and explanation in the context of regulation.

%------------------------------------------------------------------
\section{Conclusion}
\label{sec:conclusion}
%------------------------------------------------------------------

This paper presented an overview of the ongoing work in demonstrating world-leading research in mobile autonomy, focused on challenging on-road and off-road driving scenarios.
We discussed our approach for addressing the fundamental technical issues to overcome critical barriers to assurance and regulation for large-scale deployments of autonomous systems.
We foresee a future where robots can robustly sense their environment, can assess their own capabilities, and, vitally in the purpose of assurance and trust, can provide causal explanations for their own decisions.

%------------------------------------------------------------------
\section*{Acknowledgments}
%------------------------------------------------------------------

This work was supported by the Assuring Autonomy International Programme, a partnership between Lloyd's Register Foundation and the University of York.

%------------------------------------------------------------------
\bibliographystyle{IEEEtran}
\bibliography{biblio}
%------------------------------------------------------------------

\end{document}